\documentclass[12pt]{article}

\usepackage{comment}

\usepackage{times}
\usepackage{setspace}
\usepackage[margin=1in]{geometry}
\usepackage{fancyhdr}
\usepackage[numbers]{natbib} 
\usepackage{graphicx}
\usepackage{caption}
\usepackage{booktabs}
\usepackage{array}
\usepackage{float}
\usepackage{url}
\usepackage{lipsum}
\usepackage{tabularx}
\usepackage{booktabs}
\usepackage{adjustbox}
\usepackage{array}
\usepackage{comment}
\usepackage{authblk}  

\usepackage{graphicx}
\usepackage{subcaption}
\usepackage{amsmath}
\usepackage{amssymb}

\AtBeginDocument{%
  }

\begin{document}

\title{Automatic Contouring of Spinal Vertebrae on X-Ray
using a Novel Sandwich U-Net Architecture}

\author[1]{Sunil
Munthumoduku Krishna Murthy}
\author[5]{Kumar Rajamani}
\author[6]{Srividya Tirunellai Rajamani}
\author[2]{Yupei Li}
\author[2]{Qiyang Sun}
\author[1,2,3,4]{Björn W.
Schuller}

\affil[1]{CHI – Chair of Health Informatics, Technical University of Munich, Germany}
\affil[2]{GLAM – Group on Language, Audio, \& Music, Imperial College London, UK}
\affil[3]{Munich Center for Machine Learning, Germany}
\affil[4]{MDSI – Munich Data Science Institute, Germany}
\affil[5]{Marwadi University, India}
\affil[6]{University of Augsburg, Germany}

\affil[1]{\texttt{sunil.munthumoduku@tum.de};}

\date{}
\maketitle
\begin{abstract}

In spinal vertebral mobility disease, accurately extracting and contouring vertebrae is essential for assessing mobility impairments and monitoring variations during flexion-extension movements. Precise vertebral contouring plays a crucial role in surgical planning; however, this process is traditionally performed manually by radiologists or surgeons, making it labour-intensive, time-consuming, and prone to human error. In particular, mobility disease analysis requires the individual contouring of each vertebra, which is both tedious and susceptible to inconsistencies. Automated methods provide a more efficient alternative, enabling vertebra identification, segmentation, and contouring with greater accuracy and reduced time consumption. In this study, we propose a novel U-Net variation designed to accurately segment thoracic vertebrae from anteroposterior view on X-Ray images. Our proposed approach, incorporating a ``sandwich" U-Net structure with dual activation functions, achieves a 4.1\% improvement in Dice score compared to the baseline U-Net model, enhancing segmentation accuracy while ensuring reliable vertebral contour extraction.
\end{abstract}


\maketitle

\section{Introduction}
In the current work-from-home generation, prolonged sitting with poor posture is prevalent. This often leads to back pain and spinal disorders. Given the lower radiation exposure, X-ray imaging is commonly chosen for spinal assessments. Manual segmentation and contouring of vertebrae by clinicians are standard initial steps prior to further evaluation.


Medical image segmentation is one amongst the widely researched fields of Artificial Intelligence (AI) in the healthcare domain
\cite{Lin2017}, \cite{Yu2018}, \cite{Zhou2018}, \cite{Fabian2019}, \cite{rajamani2022deformable}, \cite{rajamani2022aaunet}, \cite{rajamani2023towards}, \cite{yuan2024deep}. Accurate segmentation of tissues, organs, lesions, and nodules in medical images obtained from diagnostic equipment like X-ray is crucial. Precise segmentation facilitates faster diagnosis and efficient treatment planning. However, since X-ray data is inherently two-dimensional, it provides less anatomical detail compared to advanced imaging modalities such as Computed Tomography (CT) or Magnetic Resonance Imaging (MRI). Therefore, leveraging appropriate deep learning techniques is vital to improving segmentation performance and ensuring reliable medical image analysis. Today, Convolutional Neural Networks (CNNs) play a crucial role to automate the manual effort of doctors in contouring vertebrae \cite{zhang2023customized}, \cite{shen2023dskca}, \cite{huang2020unet}, \cite{liang2023n}.U-Net based networks have been demonstrated to be well suited for the task of semantic segmentation in the
field of computer vision \cite{du2020medical},\cite{su2022improved}. 

Additionally, regulatory standards for AI in medical devices are quite stringent. Approval is granted only if the software demonstrates exceptional performance and promising accuracy, given its implications for human health. Hence, medical image segmentation is a field of active research in order to strive towards higher performance as well as accurate diagnosis.

In this work, we propose a novel U-Net architecture which uses Rectified Linear Unit (ReLU) activation functions in the first half of the ``U"  -- that is, during down-sampling -- and attention based ReLU (AReLU) in the second half of the ``U" -- that is, during up-sampling to achieve a higher Dice score. 
Our approach includes six encoder blocks that employ a ReLU activation function and five decoder blocks that employ an AReLU activation function.
Our approach will improve the accuracy on partial vertebra and edge cases, which is shown in results section.

\section{Related Work}

Segmentation of an image has been studied over decades and also considered as one of the most tedious tasks in image processing and computer vision. A central problem of image  segmentation is to segregate objects from background
\cite{jiang2004som}, \cite{minaee2021image}, \cite{zheng2021rethinking}, \cite{wu2023conditional}. Medical datasets have been extensively used for segmentation techniques to distinguish input images and retrieve adequate knowledge about the Region Of Interest (ROI) on the image \cite{sun2023boundary}. One of the tedious aspect is the identification of the neighboring tissues around organs of interest. In most of the cases, the neighboring organ tissues might have almost similar intensities or characteristics and fuzzy boundaries compared to the ROI. Hence, this task is considered as one of the complex tasks to be accomplished during segmentation of medical images \cite{sarma2021comparative}, \cite{rajamani2023novel}, \cite{huang2020wnet}.

Zhu et al.\ \cite{zhu2020tooth} propose a Mask Region-Based Convolutional Neural Network (R-CNN) for automatic teeth segmentation and detection in X-Ray images. The Mask R-CNN method is employed in both semantic segmentation and object detection. The main objective of this paper is to identify and segment teeth. They demonstrate that a Mask R-CNN performs well in complex and crowded tooth structures when it comes to segmentation. Pixel accuracy (PA) is used to assess the outcomes. To address the instance segmentation challenge, they employed Mask R-CNN convolutional networks, a type of convolutional neural network that functions as a feature extractor. In their experiment, they used a deeper Residual Network (ResNet) -- ResNet101 + Feature Pyramid Network (FPN) -- as the backbone network to extract the picture features. Instead of using ROI pooling in the Mask R-CNN, they employ the ROI Align approach. The authors demonstrate that, when using ROI Align, the approximate spatial position can be maintained. 

Su et al.\ \cite{su2019object} propose an exact Mask R-CNN, or precise Mask R-CNN
to detect objects and segment instances in Very High Resolution (VHR) remote sensing images. Bounding boxes and segmentation masks are produced for each occurrence of an object in the image using this method. A two-stage process-based Mask R-CNN framework serves as the foundation for their system \cite{hellmann2021deformable}. A Region Proposal Network (RPN) forms the first stage. A binary mask prediction branch and a Fast R-CNN classifier make up for the second stage. Instead of using ROI Align, they use precise ROI pooling which has a continuous gradient on bounding box coordinates. Further, accurate ROI Pooling does not include any quantization of coordinates and hence can identify thick and small man-made objects. 


The activation function is one of the major building blocks for many learning tasks in neural networks \cite{kunc2024exploring},\cite{jagtap2023important}, \cite{dubey2022activation},\cite{mercioni2023brief}, \cite{biswas2023non}, \cite{rasamoelina2022large}.
Element-wise activation functions influence both the expressivity power and the learning dynamics. Chen et al.\ \cite{chen2020arelu} propose a new learnable activation function which formulates an element-wise attention mechanism. Each layer is introduced with an element-wise, sign-based attention map for pre-activation of a feature map. The scaling is based on the sign. Adding the attention module with ReLU amplifies positive elements and suppresses negative elements. This makes their network more resistant to the gradient vanishing.

Rasamoelina et al.\ \cite{rasamoelina2022large} compare the 18 most frequently used activation functions on multiple datasets. Furthermore, they explore the shape of the loss landscape of those different architectures with various activation functions. They introduce a new locally quadratic activation function namely Hytana alongside one variation Parametric Hytana which benefits common activation functions and also address the dying ReLU problem.

Vijay Badrinarayanan et al.\ \cite{badrinarayanan2017segnet} propose SegNet which is a deep fully convolutional neural network designed for semantic pixel-wise segmentation. Its architecture consists of an encoder-decoder structure, where the encoder mirrors the 13 convolutional layers of VGG16, and the decoder upscales feature maps using max-pooling indices from the encoder rather than learning upsampling, making it memory-efficient.The network is optimized for scene understanding applications, balancing segmentation accuracy, computational efficiency, and memory usage.

Mingxing Tan et al.\ \cite{tan2019efficientnet}  propose a study which systematically analyses model scaling and finds that balancing network depth, width, and resolution leads to better performance. Based on this insight, the authors propose a compound scaling method that uniformly adjusts these dimensions using a simple yet effective scaling coefficient.By applying this approach, they scale up MobileNets and ResNet and further use neural architecture search to design a new baseline network, leading to the EfficientNet family.


\section{Dataset and baseline model architecture}

The dataset and base model architecture on which we conduct our experiments are detailed next.

\subsection{Burapha University's (BUU) Dataset}\label{AA}
The dataset we use for our experiments is a publicly available spine dataset collected by the faculty of Burapha University, Thailand. This dataset consists of 400 pairs of X-Ray images of different resolutions and different X-Ray manufacturers. There are two views in the dataset, the
anteroposterior view (AP View) and the lateral view (LA view) depending on the angle at which the X-Ray is acquired. 
The dataset consists of 127 male and 273 female patients. The patient age ranges from 6 years to 89 years with the average age being 50.21 years.

\subsection{Dataset Preparation}
\begin{figure}[t!]\label{Annotation_img}
    \includegraphics[width=8cm]{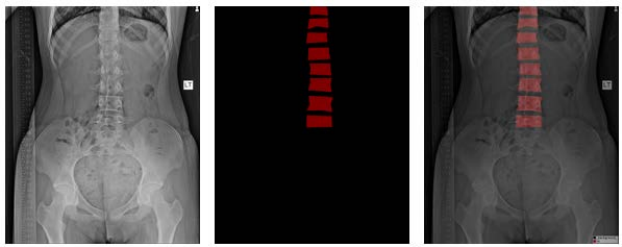}
    \centering
    \caption{
    Annotated thoracic vertebrae on X-Ray image in AP view sample for one patient:
    Original, annotated, and super-imposed image (from left to right), respectively.}
    \label{Annotation_img1}
\end{figure}

In this study, we focus on the thoracic region using AP view X-Ray images. A total of 300 AP view X-Ray images are utilised, with age variations randomly distributed across 180 female and 120 male subjects. For model development, 240 images are allocated for training and validation, with 80\% or 192 images randomly selected for training and the remaining 20\% or 48 images for validation. An additional set of 60 images are reserved for testing to assess model performance.

.
\newline
Stage 1:

All the X-Ray images in AP view are manually annotated with the help of guidance from a physician and a radiologist. The annotation is carried out via a polygon point to segment the thoracic vertebrae using the LabelMe annotation software. Fig.~\ref{Annotation_img1} visualises the original, segmented and super imposed images respectively for one patient. \newline

Stage 2: 

The augmentation pipeline incorporates a variety of geometric transformations (scaling, shifting, rotating, and flipping), color adjustments (brightness and contrast), and noise effects (blurring, downscaling, and Gaussian noise) to enhance the model's generalization. Specifically, a 512×512 region was randomly cropped from each image, with padding applied if the image dimensions were smaller than 512×512. Brightness and contrast were adjusted randomly within a range of ±0.25 with a 50\% probability. The images were randomly scaled between 85\% and 115\%, resized to 512×512 using nearest-neighbor interpolation, and subjected to random shifts of up to 32.5\% of their size. Additionally, rotation was applied within a range of ±15°. To mitigate background noise, Gaussian filtering was applied to contrast-enhanced images. The images were also downscaled to 15–25\% of their original size and subsequently upscaled to simulate low-resolution effects. Furthermore, Gaussian noise with a variance between 0.05 and 0.1 was introduced, while horizontal flipping was performed with a 25\% probability. Finally, Gaussian blur with a kernel size of 1 was applied to smooth sharp transitions and high-frequency details, effectively reducing noise.\newline

Stage 3: Re-sizing of the image data to 512*512.

\subsection{Baseline Model Architecture : U-Net}

U-Net is one of the most commonly used architecture for semantic segmentation \cite{singh2023semantic}, \cite{krithika2022review}, \cite{cao2022swin}, \cite{hatamizadeh2022unetr}, \cite{zhou2019unet++}. Semantic
segmentation is when each pixel of an image is assigned a class label. 
U-Net architecture consists of 2 important
stages. The first stage is the path called the contracting path and the second stage is the path called the expanding path.
The activation function used is a non-linear which is ReLU.
U-Net is also called as the encoder and decoder network. Since the image is down-sampled and up-sampled, it is important to use skip connections to make sure that the information from the previous
layers are not lost. One thing to note is that during the contracting path, the size of the image is
reduced due to the introduction of max-pooling to the input pixels.

\section{Proposed Methodology} 
\begin{figure}[t!]\label{architecture}
\includegraphics[width=.62\linewidth, height=2.7\textheight,keepaspectratio]
    {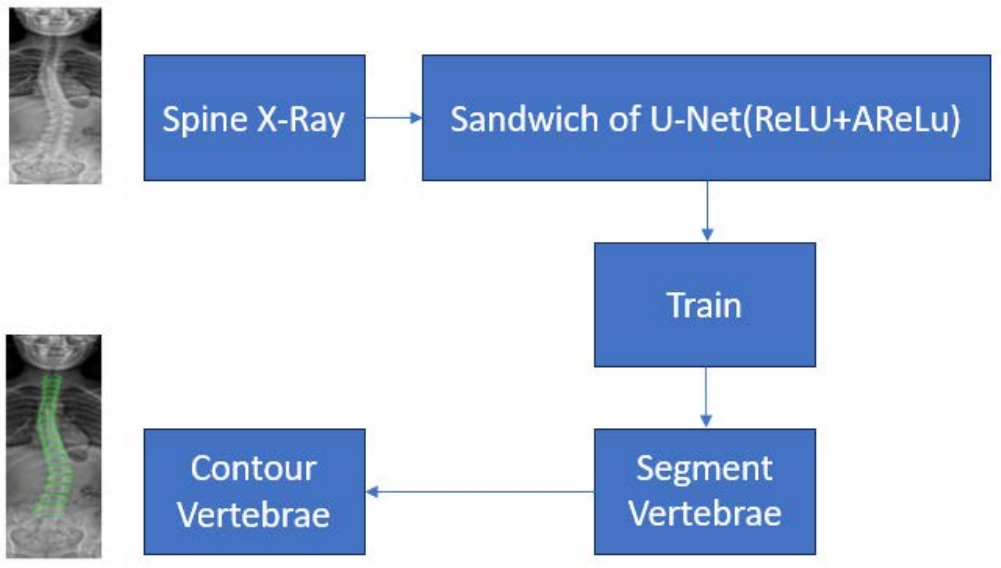}
    \centering
    \caption{Algorithm Architecture Flow.
    }
    \label{architecture1}
    
\end{figure}

We propose a novel architecture in this section, the schematic flow chart of which is shown in Fig.~\ref{architecture1}. We present our methodology in details in subsections.

\subsection{Novel Optimised Sandwich U-Net}

\begin{figure}[t!]\label{UnetModelNovel}
  \includegraphics[width=.6\linewidth, height=0.9\textheight, keepaspectratio]
    {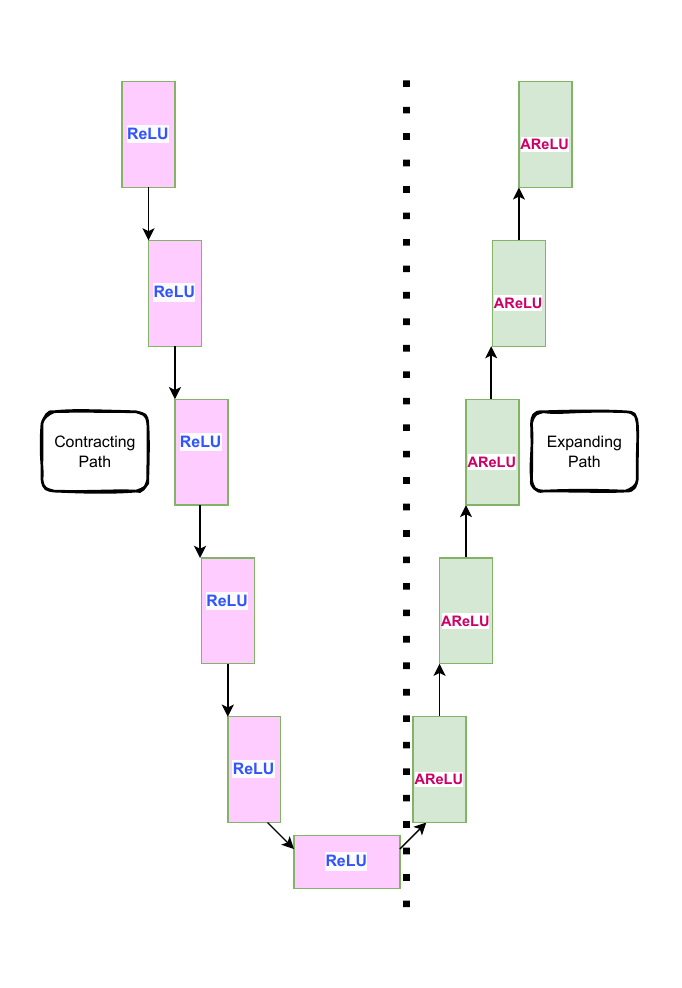}
    \centering
    \caption{Proposed Novel U-Net architecture with a sandwich of ReLU and AReLU activation functions. (a).\ ReLU activation used during the down-sampling (contracting or encoder) phase, and (b).\ AReLU activation used during the up-sampling (expanding or decoder) phase.}
    \label{UnetModelNovel1}
    
\end{figure}

In this paper, we propose a novel U-Net architecture 
In total, we incorporate
six down-sampling layers and five up-sampling layers.
With the inclusion of an additional down-sampling layer, the network allocates greater capacity to learning a richer and more comprehensive representation of the input. This enhancement enables the extraction of deeper, more abstract, and globally relevant features, thereby improving the network's ability to understand fine-grained details and prioritize feature extraction. Furthermore, the added down-sampling layer reduces spatial resolution in the bottleneck, facilitating the efficient capture of global contextual information.  We apply ReLU as activation function for every layer in the down-sampling phase (the first half of the ``U" shown in Fig.~\ref{UnetModelNovel1}) and apply the AReLU activation function for every layer in the up-sampling phase (the second half of the ``U" shown in 
Fig.~\ref{UnetModelNovel1} depicts the novel U-Net architecture.
The different activation functions during down-sampling and up-sampling allows the encoder to focus on robust feature extraction while enabling the decoder to reconstruct features adaptively, emphasizing important regions.

\begin{figure}[t!]
\includegraphics[width=0.7\linewidth,keepaspectratio]
{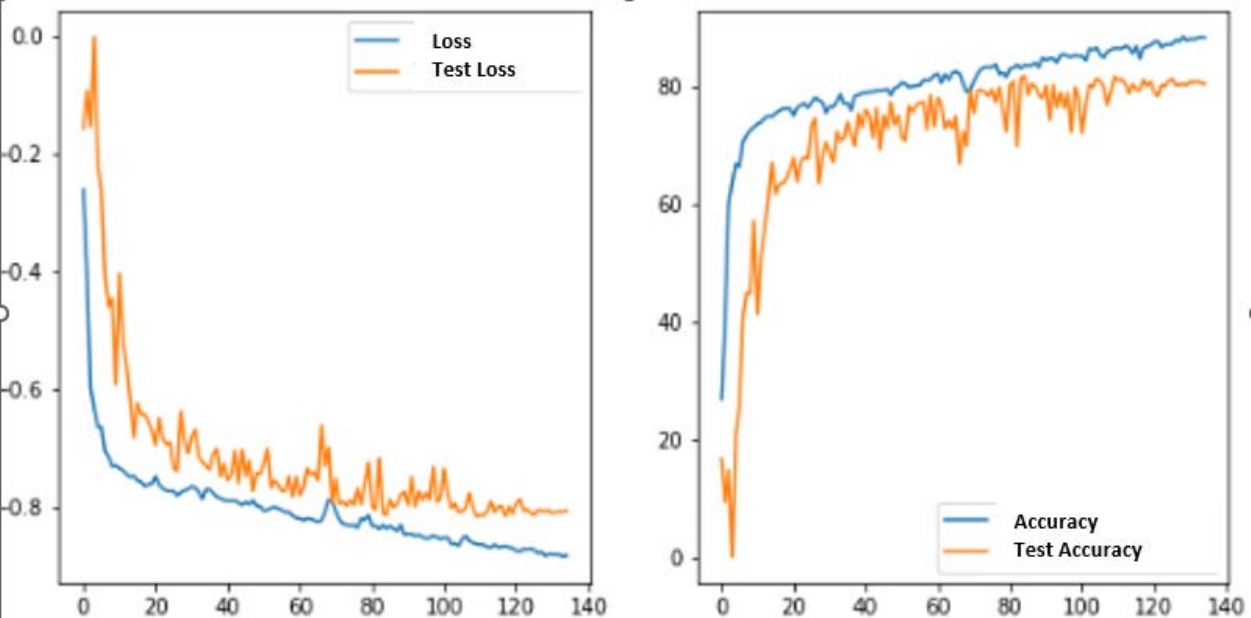}
    \centering
    \caption{Plot of the Loss and Dice Score of the U-Net model.}
    \label{lossgraph12}
   
\end{figure}

\begin{figure}[t!]
\includegraphics[width=0.7\linewidth,keepaspectratio]
{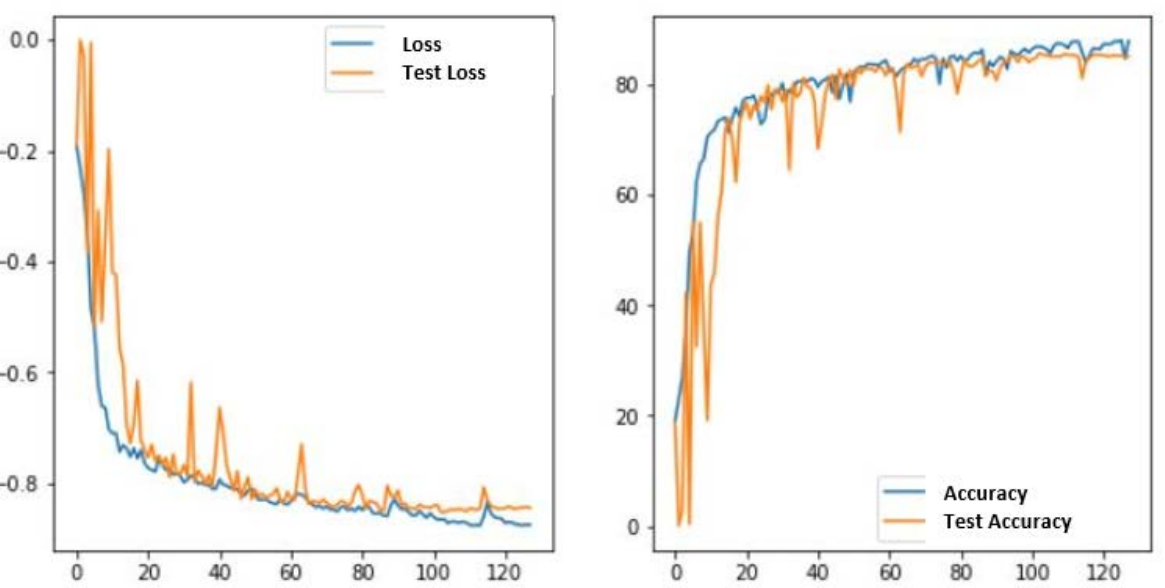}
    \centering
    \caption{Plot of the Loss and Dice Score of the optimised sandwich model.}
    \label{lossgraph1}
  
\end{figure}

\begin{figure}[t!]
\centering
    \includegraphics[width=.6\linewidth, height=0.75\textheight, keepaspectratio]
    {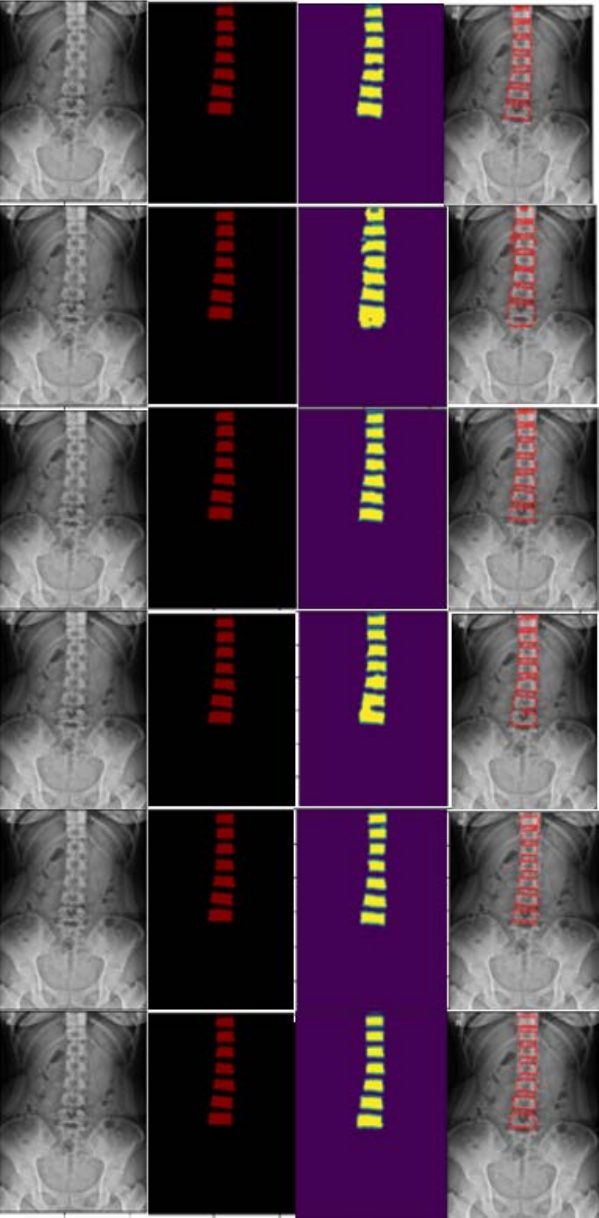}
    \centering
    \caption{Incremental addition of AReLU to each layer during up-sampling. The 6 rows, from top to bottom, show turning each layer in the up-sampling into AReLU activation, incrementally. The 4 columns, from left to right, contain   (a).\ original image, (b).\ ground truth, (c).\ predicted segmentation, and (d).\ predicted contours, respectively.}
    \label{IncrementalArelu1}
   
\end{figure}

\subsection{ReLU during down-sampling -- 1st half of the ``U"}

An activation function supports the model to use the most significant features and suppress the irrelevant features.
The ReLU activation function is a non-linear activation function that has gained significant popularity deep learning. 
ReLU does not activate all the neurons at the same time.
This is one of the significant reasons for using the ReLU function when compared to other activation functions. It avoids the vanishing gradient problem, ensures sparse activations by setting negative values to zero. This helps in focusing on essential features during the encoding process without introducing additional learnable parameters.

Here, in our implementation during the down-sampling/contracting path of the data, ReLU ($f(x)=max(0,x)$) is used as an activation function during model building. This means that the neurons of the model's layer shall be deactivated if the output of the linear transformation of a layer is less than zero.

\subsection{Attention based ReLU during up-sampling -- 2nd half of ``U"}

The AReLU activation function is used during the up-sampling or expansion path of data. AReLU enables faster model training with smaller learning rates as its a learnable activation function with a clamp and sigmoid function; it also makes the model training more resistant to the gradient vanishing.\cite{chaturvedi2022analyzing}. This mechanism dynamically adjusts the significance of each spatial and channel-wise feature map based on its relevance, thereby enhancing feature reconstruction. During the up-sampling process, where spatial details are reconstructed and encoder-decoder features are blended, the AReLU  ensures the network prioritizes salient features while suppressing irrelevant or redundant information. By applying attention across spatial dimensions or channels, it enhances the decoder's capacity to capture fine-grained details and global contextual information. This approach facilitates improved focus on critical regions, contributing to precise and effective reconstruction.

\begin{equation}
\label{eq1}
 L(x_i, \alpha, \beta) =
    \begin{cases}
      C(\alpha)x_i, \quad x_i\le 0 \\
      \sigma(\beta)x_i, \quad x_i\ge 0\\     
    \end{cases}       
\end{equation}

In Equation \ref{eq1}, $ x_i$ is the input of the current layer.  $\alpha$  and  $\beta$ are the learnable parameters. C(·) clamps the input variable into [0.01, 0.99]. $\sigma$ is the sigmoid function.

\begin{equation}
\label{eq2}
R(x_i)  =
    \begin{cases}
       0, \quad x_i\le 0\\
       x_i, \quad x_i\ge 0\\     
    \end{cases}       
\end{equation}

\begin{equation}
\label{eq3}
\begin{split}
    F(x_i, \alpha, \beta) =  R(x_i) + L(x_i, \alpha, \beta)
\end{split}
\end{equation}

\begin{equation}
\label{eq4}
F(x_i, \alpha, \beta)  =
    \begin{cases}
       C(\alpha)x_i, \quad x_i\le 0\\
       (1+\sigma(\beta))x_i, \quad x_i\ge 0\\     
    \end{cases}       
\end{equation}


 where Equation \ref{eq1} is a learnable activation using ELSA. 
Equation \ref{eq2}  is a standard rectified linear unit. Adding Equation \ref{eq1} and Equation \ref{eq3}  will result in the AReLU learnable activation function Equation \ref{eq4}  which is used during up-sampling.

In this study, we carry out a layer-by-layer assessment of the experiment, utilising the AReLU activation function during the up-sampling (decoder) phase to evaluate its effectiveness and accuracy. Table~\ref{ARELUUpSample} and Fig.~\ref{IncrementalArelu1} illustrate how the progressive application of AReLU at each layer enhances accuracy, particularly around the vertebrae edges.

Our experimental findings indicate that applying the AReLU activation function across all layers in the up-sampling path achieves the highest segmentation performance.

\begin{table}[t!]
\begin{center}
\centering
\caption{Impact of the AReLU during up-sampling.}
\label{ARELUUpSample}
\begin{tabular}{ r|r  }
\hline
\textbf{\#\,AReLU} & \textbf{\%\,Dice Score} \\
\hline
0  & 80.13 \\
\hline
1 & 80.22 \\
\hline
2 & 81.02 \\
\hline
3 & 81.48 \\
\hline
4 & 82.98 \\
\hline
\textbf{5} & \textbf{83.58} \\
\hline
\end{tabular}
\end{center}
\end{table}

\subsection{Selecting the optimal $\alpha$ (Alpha) and $\beta$ (Beta)}
The alpha parameter in the AReLU activation function introduces a small slope for negative inputs, which will make neurons inactive and stop learning. 

Beta introduces a shift or bias to better handle data distributions.
To initialise the right learnable parameters, we experimented with various combinations of alpha and beta as mentioned in  Table~\ref{learnableparamscombination}. Each combination resulted in different accuracy. However, the best accuracy was achieved with alpha initialised to 0.9 and beta to 0.9.

\begin{table}[t!]
\begin{center}
\centering
\caption{Changes in Dice Score with varying alpha and beta parameters in the AReLU.}
\label{learnableparamscombination}
\begin{tabular}{ r|r|r  }
\hline
\textbf{$\alpha$} & \textbf{$\beta$} & \textbf{\%\,Dice Score} \\
\hline
0.75 & 1.5 & 82.4 \\
\hline
0.75 & 2.0 & 83.0 \\
\hline
\textbf{0.90} & \textbf{0.9} & \textbf{83.6} \\ 
\hline
0.99 & 1.5 & 81.8 \\
\hline
0.99 & 2.0 & 82.3 \\
\hline
\end{tabular}
\end{center}
\end{table}

\section{Experimental Results}

\subsection{Development Environment }

In this study, model is trained for vertebra segmentation on google co-lab environment using Python III Google Compute Engine back-end Graphics Processing Unit (GPU) with Random Access Memory (RAM) 12.6\,GB and a disk space of 80\,GB. 

\subsection{Training and Results of the U-Net model}
The loss function and performance metric, as shown in Fig.~\ref{lossgraph12}, were assessed using Dice Loss and the Dice Coefficient, respectively. The Adaptive Moment Estimation (ADAM) optimiser was selected due to its adaptive learning rates and efficient gradient updates, which help stabilize convergence. The ReLU activation function was employed to introduce non-linearity while preventing the vanishing gradient problem. Early stopping was implemented to monitor validation loss, preventing over-fitting and ensuring optimal generalisation. Although the model was initially set to train for 200 epochs, early stopping halted training at 140 epochs when no further improvement was observed in validation loss. 
The training results showed a Dice score of 0.9017, while the test results on 60 test dataset showed a Dice score of 0.8013.
The contours predicted by this model are not very precise. It fails to accurately contour the borders of vertebrae, particularly in delineating the borders of the vertebrae, as illustrated in Fig.~\ref{Unet_And_Novel1}.

\subsection{Training the Parameters of the Novel Optimised Sandwich U-Net}

As illustrated in Fig.~\ref{lossgraph1}, Dice Loss was used as the loss function, while the Dice Coefficient served as the performance metric to evaluate segmentation accuracy. The ADAM optimizer was selected due to its ability to adapt learning rates for each parameter, improving convergence stability. The ReLU activation function was applied in the down-sampling layers to introduce non-linearity and mitigate the vanishing gradient problem. In the up-sampling layers, the AReLU activation function was utilized, with alpha and beta values initialized to 0.9, allowing dynamic adaptation of activation thresholds. Early stopping was implemented to monitor validation loss and prevent over-fitting. Although the model was initially set to train for 200 epochs, early stopping terminated training at 120 epochs when no further improvement was observed in validation loss.

\subsection{Results and Discussion of the Novel Optimised Sandwich U-Net}
Our innovative sandwich of combining two different activation functions during the first and second half of the ``U" yields a promising Dice score compared to the conventional model.
The training results show a Dice score of 0.9117. The test results on 60 test dataset show a Dice score of 0.8358 which is notably higher than the conventional model, demonstrating superior segmentation accuracy on unseen data.

\begin{figure}[t!]\
\centering
    \includegraphics[width=\linewidth]
    {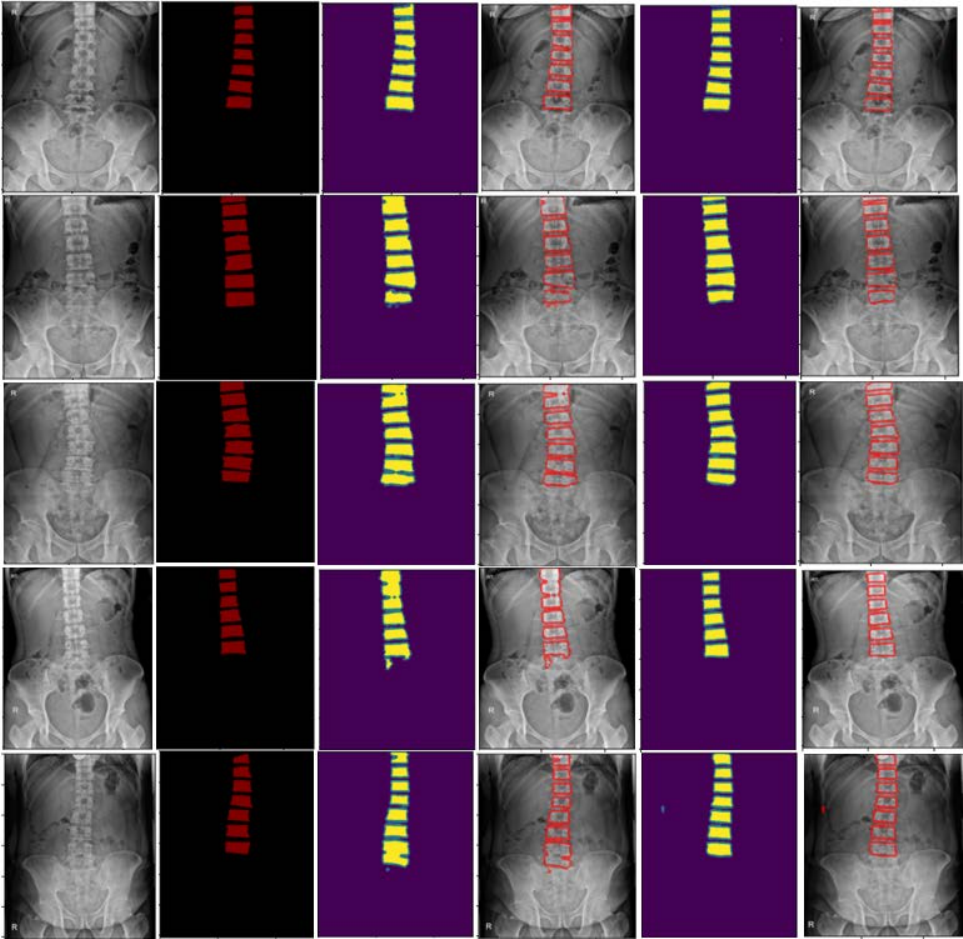}
    \centering
    \caption{Predicted contours by the U-Net model and the novel sandwich U-Net model. The 6 columns, from left to right, contain   (a).\ original image, (b).\ ground truth, (c).\ predicted segmentation using the U-Net model, (d).\ predicted contours using the U-Net Model, (e).\ predicted segmentation using our novel sandwich U-Net model, and (f).\ predicted contours using our novel sandwich U-Net model, respectively.}
    \label{Unet_And_Novel1}

\end{figure}

\begin{figure}[t!]\label{contours}
    \includegraphics[width=0.3\linewidth, height=1.9\textheight, keepaspectratio]{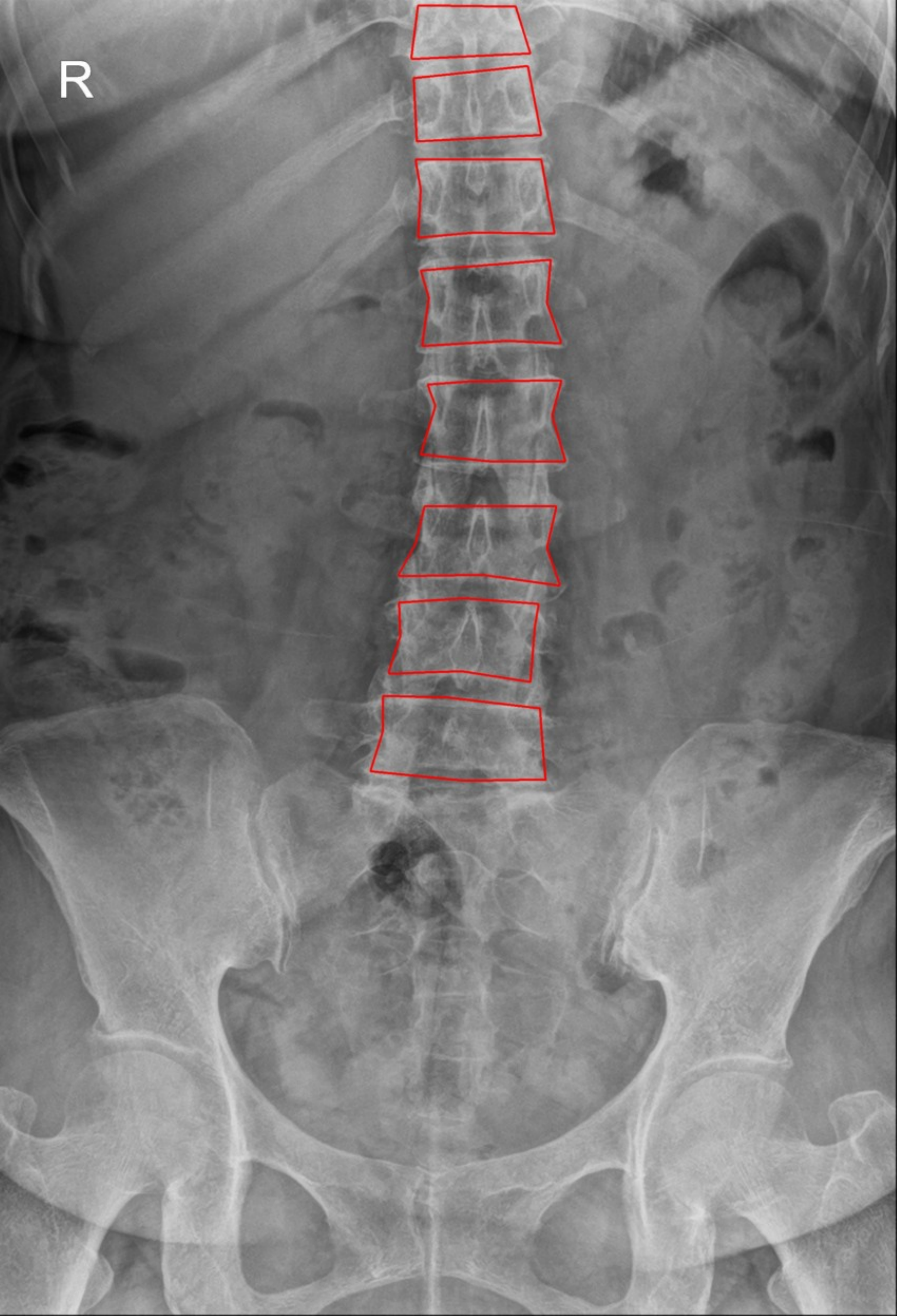}
    \centering
    \caption{Contours predicted on an unseen data using Proposed Novel U-Net Architecture.}
    \label{contours1}
   
\end{figure}

Fig.~\ref{Unet_And_Novel1} and Fig.~\ref{contours1} illustrates that the proposed model produces segmentation contours that closely correspond to the ground truth, particularly along the boundaries of the lower vertebral regions, where precise delineation is essential. Compared to the conventional model, the proposed approach exhibits enhanced edge preservation and structural consistency, likely due to the adaptive application of dual activation functions. In contrast, the conventional model struggles with capturing finer details, resulting in smoother but less accurate boundaries. This contrast suggests that the proposed model more effectively represents complex vertebral structures while minimizing misclassification errors, leading to more reliable segmentation results.

A detailed examination of the predicted vertebra segmentation further highlights that the proposed model successfully retains structural details, especially in complex boundary areas. However, the conventional model shows a higher incidence of segmentation errors, particularly in the lower vertebrae, where anatomical variations are more significant. These errors primarily appear as boundary misalignments, often causing over-segmentation or under-segmentation. The findings indicate that leveraging dual activation functions enhances feature extraction, thereby improving segmentation accuracy and reducing contour irregularities and misclassification.

\subsection{Comparison of Models}
The performance of the proposed Novel U-Net Architecture was evaluated against established deep learning-based segmentation methods, including SegNet, EfficientNet, and U-Net. The comparative analysis, presented in the table Table~\ref{modelcomparison} , highlights the Dice score achieved by each method.

We conduct a statistical analysis using the paired t-test to evaluate the significance of differences in the results presented in Table~\ref{modelcomparison}. Specifically, we perform three key comparisons: (1) SegNet vs Proposed Novel U-Net Architecture, (2) EfficientNet vs Proposed Novel U-Net Architecture, and (3) U-Net vs Proposed Novel U-Net Architecture. The p-values reported in Table~\ref{statscomparison} indicate that the differences in evaluation result is statistically significant, confirming that the proposed model demonstrates meaningful improvements over existing architectures.

\begin{table}[t!]
\begin{center}
\centering
\caption{Dice Score comparison between the base model and the novel model.}
\label{modelcomparison}
\begin{tabular}{ l|r }
\hline
\textbf{Model} & \textbf{\%\,Dice Score} \\
\hline
SegNet & 77.20 \\
\hline
EfficientNet & 79.46 \\
\hline
U-Net & 80.13 \\
\hline
 Proposed Novel U-Net Architecture & 83.58 \\
\hline
\end{tabular}
\end{center}
\end{table}

\begin{table}[t!]
\begin{center}
\centering
\caption{Statistical analysis using paired t-test. P-values are represented.}
\label{statscomparison}
\begin{tabular}{ l|r }
\hline
\textbf{Model} & \textbf{P-Values for Dice Score} \\
\hline
SegNet vs  Proposed Novel U-Net Architecture & $<$ 0.001 \\
\hline
EfficientNet vs Proposed Novel U-Net Architecture & $<$ 0.001 \\
\hline
U-Net vs Proposed Novel U-Net Architecture & $<$ 0.001  \\
\hline
\end{tabular}
\end{center}
\end{table}

Fig.~\ref{Unet_And_Novel1} illustrates the accuracy of the predicted contours in both the U-Net and the novel sandwich U-Net models.

\section{Discussion, Conclusion and Future Directions}


In this study, we optimised the U-Net model by employing ReLU activation in the down-sampling (contracting) path and AReLU activation in the up-sampling (expansive) path, with clamp of alpha and beta parameters. The segmentation predictions generated by the optimised model enabled precise contouring of thoracic vertebrae in AP view, demonstrating visible improvements over the conventional approach.

The integration of dual activation functions (ReLU and AReLU) resulted in a Dice score of 83.58\%, representing a 4.1\% improvement over the standard U-Net for vertebrae contouring. The model effectively delineated partial vertebrae and vertebral borders with enhanced accuracy. Multiple experiments were conducted, evaluating different pooling layers and loss functions, with the highest Dice score of 83.58\% achieved after training for 200 epochs using the Dice loss function.

The extracted vertebra contours in the thoracic region are crucial for various healthcare applications, including spinal deformity assessment, fracture detection, and preoperative planning for surgeries such as spinal fusion. These contours facilitate automated measurements of vertebral alignment, curvature, and spacing, aiding in the early diagnosis of conditions like scoliosis and kyphosis. The Dice Score of 83.58\% is justified given the inherent challenges of thoracic X-Ray imaging, such as overlapping anatomical structures, variations in patient positioning, and low contrast in certain regions. While a higher Dice Score would indicate even finer segmentation accuracy, the achieved performance provides a reliable tool for assisting radiologists and surgeons in decision-making. In clinical practice, segmentation models with this level of accuracy significantly reduce the time and effort required for manual annotation, offering an effective solution for enhancing diagnostic workflows. 

For future work, enhancing model accuracy will necessitate access to larger, well-annotated datasets and additional computational resources. Despite these challenges, the current model demonstrates practical effectiveness in medical imaging. Moving forward, we intend to explore advanced training strategies for deeper and more complex architectures to further improve segmentation performance. Furthermore, we plan to validate our approach on publicly available medical image datasets to evaluate its robustness, applicability, and generalisability across diverse clinical scenarios.

\section{Limitation}
This study has certain limitations. Since the dataset was manually annotated and specifically customized for the thoracic region, direct comparisons with recent state-of-the-art segmentation methods were not feasible. Consequently, we have limited our comparisons to SegNet, EfficientNet and U-Net. Additionally, hardware constraints restricted both the batch size and image resolution during training, which may have impacted model performance. The Dice score could potentially be improved by increasing the batch size with a more advanced hardware configuration. Furthermore, this study focuses exclusively on the thoracic region and the AP view of spine X-rays, limiting its generalisability to other anatomical regions and imaging perspectives.

\bibliographystyle{ACM-Reference-Format}
\bibliography{references}

\end{document}